\documentclass[aps,prd,%multicol,twocolumn, 
preprintnumbers,groupedaddress]{revtex4}
\usepackage{epsbox}
\usepackage{amsmath,amssymb}
\usepackage{bm}
\usepackage[dvips]{graphicx}

\begin{document}
\preprint{\vtop{
{\hbox{YITP-16-123}\vskip-0pt
%                 \hbox{KANAZAWA-11-03} \vskip-0pt
%                 \hbox{arXiv:????.????[hep-ph]} 
}
}
}

\date{\today}

\title{ 
Tetra-Quark Interpretation of $\bm{X(3872)}$ and $\bm{Z_c(3900)}$ Revisited 
}

\author{
Kunihiko Terasaki   %authors' name%
}
\affiliation{
Yukawa Institute for Theoretical Physics, Kyoto University,
Kyoto 606-8502, Japan }
%\\    Institute for Theoretical Physics, Kanazawa University, 
%Kanazawa 920-1192, Japan}

\begin{abstract}
{
In relation to the newly observed bottom-strange $X(5568)^\pm$ mesons, we revisit 
our tetra-quark interpretation of $X(3872)$ and $Z_c(3900)$. 
It is discussed that our assignment of $X(5568)^\pm$ to charged components of 
iso-triplet bottom partners of $D_{s0}^+(2317)$ is compatible with the revised version 
of our tetra-quark interpretation of $X(3872)$ and $Z_c(3900)$. 
}
\end{abstract}

\maketitle

After the discovery of $D_{s0}^+(2317)$~\cite{D_{s0}-exp,Belle-D_{s0}}, observations of 
many heavy mesons have been reported~\cite{PDG16}, and various interpretations 
of them have been proposed~\cite{Swanson}. 
In addition, recently, charged (iso-triplet) bottom-strange mesons, $X(5568)^\pm$, 
have been observed~\cite{strange-bottom-D0}, though not 
confirmed~\cite{strange-bottom-LHCb} yet, and then, they have been 
interpreted~\cite{strange-bottom-th} as bottom-partners of $D_{s0}^+(2317)$. 
In relation to the newly observed $X(5568)^\pm$, we revisit our tetra-quark 
interpretation of $X(3872)$ and its  partners $Z^{\pm,0}_c(3900)$~\cite{Z_c(3900)-ex}  
(or $X(3900)$~\cite{PDG16}) with an opposite charge-conjugation ($\mathcal{C}$) 
property. 

Tetra-quark states are classified into the following four groups,  
%%%%%%%%%%%%%%%%%%%%%%%%%%%%%%%%%%%%%%%%%%%%%%%%%%%%%%%%%%%%%%%%%%%%%%%%
\begin{eqnarray} 
&&\hspace{-8mm} \{qq\bar q\bar q\} =  
[qq][\bar q\bar q] \oplus (qq)(\bar q\bar q)  
\oplus \{[qq](\bar q\bar q)\oplus (qq)[\bar q\bar q]\}, 
                                                   \label{eq:4-quark} 
\end{eqnarray} 
%%%%%%%%%%%%%%%%%%%%%%%%%%%%%%%%%%%%%%%%%%%%%%%%%%%%%%%%%%%%%%%%%%%%%%%%
in the framework of $q = u,d,s$~\cite{Jaffe} (and $c$~\cite{D_{s0}-KT}), where 
parentheses and square brackets in the above equation imply symmetry and 
anti-symmetry, respectively, of flavor wavefunctions (wfs.) under exchange of flavors 
between them. 
Each term in the right-hand-side of Eq.~(\ref{eq:4-quark}) is again classified into two 
groups with $\bar{\bm 3}_c\times {\bm 3}_c$ and ${\bm 6}_c\times \bar{\bm 6}_c$ 
of the color $SU_c(3)$. 
Although a mixing between two states with $\bar{\bm 3}_c\times {\bm 3}_c$ and 
${\bm 6}_c\times \bar{\bm 6}_c$ (which consist of quarks with common flavors and 
have the same quantum numbers) was considered at the scale of light meson 
mass~\cite{Jaffe}, such a mixing is neglected at the the scale of heavy meson mass 
under consideration. 
Next, we take states with $\bar{\bm 3}_c\times {\bm 3}_c$ as the lower lying 
ones~\cite{D_{s0}-KT} and those with ${\bm 6}_c\times \bar{\bm 6}_c$ as the higher 
ones, because a force between two quarks is attractive when the two quark state is 
of $\bar{\bm 3}_c$, $\{qq\}_{\bar{3}_c}$, while repulsive when of ${\bm 6}_c$, 
$\{qq\}_{6_c}$~\cite{color}.  
Regarding spin ($J$) of $[qq]$ and $(qq)$, its values are $J = 0$ and $1$ for 
$[qq]_{\bar{3}_c}$ and $(qq)_{\bar{3}_c}$, and $1$ and $0$ for $[qq]_{{6}_c}$ and 
$(qq)_{{6}_c}$, respectively, in the flavor symmetry limit, for the reson that their wfs. 
should be totally anti-symmetric in the limit.  
However, one might worry about large breakings of the flavor $SU_f(3)$ and $SU_f(4)$ 
symmetries, as in meson masses. 
Nevertheless, such breakings are not necessarily serious in wfs., as seen below.  
A matrix element of flavor charge is given by a related form factor of vector current 
at zero momentum-transfer squared, and the form factor is normalized to be unity in 
the flavor symmetry limit. 
This implies that its deviation from unity provides a measure of flavor symmetry 
breaking under consideration. 
Their phenomenological and measured values have been summarized in \cite{PDG96} 
as follows. 
The  form factor of strangeness-changing vector current taken between $\langle\pi|$ 
and $|K\rangle$ has been given by 
{\small $f_+^{(\pi K)}(0)$} $= 0.961 \pm 0.008$~\cite{Leutwyler}. 
This implies that the flavor $SU_f(3)$ symmetry works well in wfs. 
The form factors of charm-changing currents between $\langle \bar{K}|$ (or 
$\langle \pi|$) and $|D\rangle$ have been provided as 
%%%%%%%%%%%%%%%%%%%%%%%%%%%%%%%%%%%%%%%%%%%%%%%%%%%%%%%%%%%%%%%%%%%%%%%%
{\small $f_+^{(\bar{K}D)}(0)$}$= 0.74\pm 0.03$~\cite{PDG96}, 
{\small $f_+^{(\pi D)}(0)/f_+^{(\bar{K}D)}(0)$} 
$= 1.00\pm 0.11 \pm 0.02$~\cite{FNAL-E687} 
and {$0.99 \pm 0.08$}~\cite{CLEO-FF}. 
%%%%%%%%%%%%%%%%%%%%%%%%%%%%%%%%%%%%%%%%%%%%%%%%%%%%%%%%%%%%%%%%%%%%%%%%
From the above results, it is seen that the $SU_f(3)$ symmetry works well even in 
the world including charm mesons, and the $SU_f(4)$ symmetry breaking is not very 
serious in wfs.  
In this way, we take the above values of spin of $[qq]$ and $(qq)$ in the flavor 
symmetry limit~\cite{KT-MacKellar}. 
As the result, spin and parity ($P$) of $[qq][\bar{q}\bar{q}]$ and 
$[qq](\bar{q}\bar{q})\oplus (qq)[\bar{q}\bar{q}]$ mesons are taken to be $J^P = 0^+$ 
and $1^+$, respectively, when they are of $\bar{\bm 3}_c\times {\bm 3}_c$, while 
$J^P = (0^+,1^+,2^+)$ and $1^+$, respectively, when of 
${\bm 6 }_c\times \bar{\bm 6}_c$.  
Here, it should be noted that axial-vector $[qq][\bar{q}\bar{q}]$ states with 
$\bar{\bm 3}_c\times {\bm 3}_c$ disappear in the flavor symmetry limit, while our 
axial-vector $[qq](\bar{q}\bar{q})\oplus (qq)[\bar{q}\bar{q}]$ states survive even in the 
limit. 
However, $(qq)(\bar{q}\bar{q})$ states are not considered in this short note, because 
existence of strange scalar mesons with the isospin $|{\bm I}| = 3/2$ which can be 
given by $(nn)(\bar{s}\bar{n})$ with $n = (u,d)$~\cite{Jaffe} is not established 
yet~\cite{K-pi-3/2}. 

In our earier works~\cite{Terasaki-X,omega-rho-KT}, the flavorless axial-vector meson, 
$X(3872)$, was assigned to 
%%%%%%%%%%%%%%%%%%%%%%%%%%%%%%%%%%%%%%%%%%%%%%%%%%%%%%%%%%%%%%%%%%%%%%
{\small 
$X(+) \sim \{[cn]_{\bar{3}_c}^{1_s}(\bar{c}\bar{n})_{3_c}^{3_s} 
+ (cn)_{\bar{3}_c}^{3_s}[\bar{c}\bar{n}]_{3_c}^{1_s}\}_{1_c}^{3_s}|_{|\bm{I}| = 0},\,(n = u,d)$} 
%%%%%%%%%%%%%%%%%%%%%%%%%%%%%%%%%%%%%%%%%%%%%%%%%%%%%%%%%%%%%%%%%%%%%%
with $\bar{\bm 3}_c\times {\bm 3}_c$ as the lowest iso-singlet hidden-charm 
axial-vector tetra-quark meson, where $1_s$ and $3_s$ mean the spin-singlet 
and spin-triplet, respectively. 
However, its measured mass seems to be too high (higher by about $1600$ MeV than  
that of $D_{s0}^+(2317)$ which has been assigned to the scalar 
%%%%%%%%%%%%%%%%%%%%%%%%%%%%%%%%%%%%%%%%%%%%%%%%%%%%%%%%%%%%%%%%%%%
{\small $\hat{F}_I^+ \sim \{[cn]_{\bar{3}_c}^{1_s}
                           [\bar{s}\bar{n}]_{3_c}^{1_s}\}_{1_c}^{1_s}|^{+}_{|{\bm I}| = 1},\,(n = u,d)$} 
%%%%%%%%%%%%%%%%%%%%%%%%%%%%%%%%%%%%%%%%%%%%%%%%%%%%%%%%%%%%%%%%%%%
with the same $\bar{\bm 3}_c\times{\bm 3}_c$~\cite{D_{s0}-KT}). 
In addition, existence of iso-triplet hidden-charm scalar mesons 
%%%%%%%%%%%%%%%%%%%%%%%%%%%%%%%%%%%%%%%%%%%%%%%%%%%%%%%%%%%%%%%%%%%
{\small $\hat{\delta}^c 
\sim \{[cn]_{\bar{3}_c}^{1_s}[\bar{c}\bar{n}]_{3_c}^{1_s}\}_{1_c}^{1_s}|_{|\bm{I}| = 1},\,(n = u,d)$} 
%%%%%%%%%%%%%%%%%%%%%%%%%%%%%%%%%%%%%%%%%%%%%%%%%%%%%%%%%%%%%%%%%%%
with $\bar{\bm 3}_c\times{\bm 3}_c$ was predicted in our tetra-quark 
model~\cite{hidden-charm-scalar-KT}, and their mass was very crudely estimated to be 
$m_{\hat{\delta}^c}\simeq 3.3$ GeV by using a simple quark counting with the mass 
difference 
%%%%%%%%%%%%%%%%%%%%%%%%%%%%%%%%%%%%%%%%%%%%%%%%%%%%%%%%%%%%%%%%%%%%%%%%
$\Delta_{cs} = m_c - m_s \simeq m_{\eta_c} - m_{D_s^+} \simeq 1.0$ GeV 
%%%%%%%%%%%%%%%%%%%%%%%%%%%%%%%%%%%%%%%%%%%%%%%%%%%%%%%%%%%%%%%%%%%%%%%%
at the scale of charm meson mass and taking the mass $\simeq 2.3$ GeV of 
${D_{s0}^+(2317)}$ as the input data, independently of the observation of an indication 
of $\eta\pi^0$ peak around 3.2 GeV at the Belle~\cite{hidden-charm-scalar-Belle}. 
(The $\eta\pi^0$ peak is now named as $\hat{\delta}^{c0}(3200)$ for later convenience, 
and assigned to $\hat{\delta}^{c0}$.) 
Our quark counting seems to work, though still crude, because the predicted 
$m_{\hat{\delta}^c} \simeq 3.3$ GeV reproduces cosiderably well the measured 
$m_{\hat{\delta}^c(3200)}\sim 3.2$ GeV, as seen above. 
However, the mass of  $\hat{\delta}^{c0}$ is much lower than that of $X(3872)$, in 
spite that their constituents have common flavors. 
This suggests that these two meson states have different structure with respect to 
the color degree of freedom, i.e., $\hat{\delta}^c$ with $\bar{\bm 3}_c\times {\bm 3}_c$ 
and $X(3872)$ with ${\bm 6}_c\times\bar{\bm 6}_c$, and therefore, the mass difference 
($\Delta_{\rm color}$) between two states with ${\bm 6}_c\times\bar{\bm 6}_c$ and 
$\bar{\bm 3}_c\times{\bm 3}_c$ (whose constituent quarks have common flavors) is 
taken as 
%%%%%%%%%%%%%%%%%%%%%%%%%%%%%%%%%%%%%%%%%%%%%%%%%%%%%%%%%%%%%%%%%%%%%%%%
$\Delta_{\rm color}\simeq m_{X(3872)} - m_{\hat{\delta}^c}\simeq 0.6$ GeV 
%%%%%%%%%%%%%%%%%%%%%%%%%%%%%%%%%%%%%%%%%%%%%%%%%%%%%%%%%%%%%%%%%%%%%%%%
in this note. 
(Mass differences arising from different structures of spin and flavor wfs. are 
suspected  to be not very large at the scale of heavy meson mass.) 
Here, it should be noted that the estimated mass $m_{\hat{\delta}^c}\simeq 3.3$ GeV 
of the lowest hidden-charm scalar meson in our model (and the measured 
$m_{\hat{\delta}^c(3200)} \sim 3.2$ GeV of its candidate) are much lower than the 
mass $\simeq 3.7$ GeV of the lowest hidden-charm scalar meson in the diquark 
model~\cite{D_{s0}-Maiani} and in the unitarized chiral 
model~\cite{hidden-charm-scalar-Oset}. 
Thus, the hidden-charm scalar meson might be a clue to select a realistic model of 
multi-quark mesons, and therefore, confirmation of $\hat{\delta}^{c0}(3200)$ is awaited. 
\vspace{-5mm}
%%%%%%%%%%%%%%%%%%%%%%%%%%%%%%%%%%%%%%%%%%%%%%%%%%%%%%%%%%%%%%%%%%%%%%%%
\begin{center} 
\begin{table}[t]       
\begin{quote}
%\caption{
Table~I. Open- and hidden-charm scalar tetra-quark mesons and their flavor 
wavefunctions, where $C$, $S$ and $|\bm{I}|$ denote charm, strangeness and isospin 
quantum numbers, respectively. 
Their masses are estimated by using a quark counting, as discussed in the text. 
Tetra-quark states with $\ast$ are of ${\bm 6}_c\times\bar{\bm 6}_c$, 
and $J/\psi$ is written as $\psi$. Notations of flavor wfs. whose overall normalization 
factors are dropped are explained in the text. 
\end{quote} \vspace{2mm}
%%%%%%%%%%%%%%%%%%%%%%%%%%%%%%%%%%%%%%%%%%%%%%%%%%%%%%%%%%%%%%%%%%%%%%%%
\begin{tabular}{|c|c|c|c|c|c|c|}
\hline
$C$ & $S$ & $|\bm{I}| = 1$ 
& $|\bm{I}| = 1/2$ & $|\bm{I}| = 0$ & Mass
& 
\begin{tabular}{l}
Candidate or \\
possible %\\
decay
\end{tabular}
\\
\hline
 & &  
$\left.\begin{tabular}{l}
\vspace{-4mm}\\
\hspace{0mm}$\hat{F}_I
\sim \{[cn]_{\bar{3}_c}^{1_s}[\bar{s}\bar{n}]_{3_c}^{1_s}\}_{1_c}^{1_s}\bigl|_{|\bm{I}| =1}$ 
\vspace{-2.5mm}\\
-----------------------------------\vspace{-2mm}\\
\hspace{-1mm}$\hat{F}_I^{\ast}
\sim \{[cn]_{6_c}^{3_s}[\bar{s}\bar{n}]_{\bar{6}_c}^{3_s}\}_{1_c}^{1_s}\bigl|_{|\bm{I}| =1}$
\end{tabular}\right. 
$  \hspace{-3mm}
&  &  &
\begin{tabular}{l}
 $\sim 2.3$ GeV ($\dagger$)\vspace{-2mm}\\
 ---------------\vspace{-2mm}\\
$\sim 2.9$ GeV
\end{tabular}
& \begin{tabular}{c}
$D^+_{s0}(2317)$ \vspace{-2mm}\\
--------------------\vspace{-2mm}\\
\,$\rightarrow D_s^+\pi^0,\,DK$
\end{tabular} 
\\
\cline{3-7}
 & 
\begin{tabular}{c}
\vspace{-13mm} \\
1 
\\
\end{tabular} 
  &  & & 
$\left.\begin{tabular}{l}
\vspace{-4mm}\\
\hspace{0.5mm}$\hat{F}_0^+
\sim \{[cn]_{\bar{3}_c}^{1_s}[\bar{s}\bar{n}]_{3_c}^{1_s}\}_{1_c}^{1_s}\bigl|_{|\bm{I}| =0}$ 
\vspace{-2.5mm}\\
------------------------------------\vspace{-2mm}\\
\hspace{-1mm}$\hat{F}_0^{\ast +}
\sim \{[cn]_{6_c}^{3_s}[\bar{s}\bar{n}]_{\bar{6}_c}^{3_s}\}_{1_c}^{1_s}\bigl|_{|\bm{I}| =0}$
\end{tabular}\right.$\hspace{-1mm}  
&  \hspace{-5mm}
\begin{tabular}{l}
 $\sim 2.3$ GeV \vspace{-2mm}\\
 ---------------\vspace{-2mm}\\
$\sim 2.9$ GeV
\end{tabular}
& \hspace{0mm}
\begin{tabular}{l}
$\rightarrow  D_s^{*+}\gamma$\vspace{-2mm}\\
------------------\vspace{-2mm}\\
$\rightarrow DK,\,D_s^+\eta$
\end{tabular}
\\
\cline{2-7}
\begin{tabular}{c}
\vspace{-10mm} \\
1 
\\
\end{tabular} 
& & 
 & 
\,$\left.\begin{tabular}{l}
\vspace{-4mm}\\
\hspace{0.5mm}$\hat{D} \sim \{[cn]_{\bar{3}_c}^{1_s}[\bar{u}\bar{d}]^{1_s}_{3_c}\}_{1_c}^{1_s}$ 
\vspace{-2.5mm}\\
-----------------------------\vspace{-2mm}\\
\hspace{-1mm}$\hat{D}^{\ast} \sim \{[cn]_{6_c}^{3_s}
                                                  [\bar{u}\bar{d}]_{\bar{6}_c}^{3_s}\}_{1_c}^{1_s}$
\end{tabular}\right.$    
& & \hspace{-5mm}
\begin{tabular}{l}
$\sim 2.2$ GeV \vspace{-2mm}\\
 ---------------\vspace{-2mm}\\
$\sim 2.8$ GeV
\end{tabular}
 & \hspace{-11.5mm}$\rightarrow D\pi$
\\
\cline{3-7} 
& 
\begin{tabular}{c}
\vspace{-13mm} \\
0 
\\
\end{tabular} 
& & 
$\left.\begin{tabular}{l}
\vspace{-4mm}\\
\hspace{-1.5mm}$\hat{D}^s\sim \{[cs]_{\bar{3}_c}^{1_s}
                                              [\bar{n}\bar{s}]_{3_c}^{1_s}\}_{1_c}^{1_s}$ 
\vspace{-2.5mm}\\
\hspace{-3.5mm}
-----------------------------\vspace{-2mm}\\
\hspace{-2.5mm}$\hat{D}^{s\ast}\sim \{[cs]_{6_c}^{3_s}
                                                     [\bar{n}\bar{s}]_{\bar{6}_c}^{3_s}\}_{1_c}^{1_s}$
\end{tabular}\right.$\hspace{-5mm}    
& &  \hspace{-5mm}
\begin{tabular}{l}
 $\sim 2.4$ GeV \vspace{-2mm}\\
 ---------------\vspace{-2mm}\\
$\sim 3.0$ GeV
\end{tabular}
 & \hspace{-0mm}
\begin{tabular}{l}
$\rightarrow D^*\gamma,\,D\eta$\vspace{-2.5mm}\\
--------------------\vspace{-2mm}\\
$\rightarrow D\eta,\,D_s^+\bar{K}$
\end{tabular}
\\
\cline{2-7}
& -1 &  &  & \hspace{-5mm}
$\left.\begin{tabular}{l}
\vspace{-4mm}\\
\hspace{1.5mm}$\hat{E}^{0}\sim \{[cs]_{\bar{3}_c}^{1_s}
[\bar{u}\bar{d}]_{3_c}^{1_s}\}_{1_c}^{1_s}$
\vspace{-2.5mm}\\
-----------------------------\vspace{-1.5mm}\\
\hspace{-0mm}$\hat{E}^{\ast 0}\sim \{[cs]_{6_c}^{3_s}
[\bar{u}\bar{d}]_{\bar{6}_c}^{3_s}\}_{1_c}^{1_s}$
\end{tabular}\right.$ 
&  \hspace{-5mm}
\begin{tabular}{l}
 $\sim 2.3$ GeV \vspace{-2mm}\\
 ---------------\vspace{-2mm}\\
$\sim 2.9$ GeV
\end{tabular}
 & 
\begin{tabular}{l}
weak decay \vspace{-2mm}\\
--------------------\vspace{-2mm}\\
\hspace{-0mm}$\rightarrow D\bar{K}$
\end{tabular}
\\
\hline
 &  & 
$\left.\begin{tabular}{l}
\vspace{-4mm}\\
\hspace{0.5mm}$\hat{\delta}^c
\sim \{[cn]_{\bar{3}_c}^{1_s}
        [\bar{c}\bar{n}]_{3_c}^{1_s}\}_{1_c}^{1_s}\bigl|_{|\bm{I}| = 1}$\vspace{-2.5mm}\\
-----------------------------------\vspace{-1.5mm}\\
\hspace{-1mm}$\hat{\delta}^{c\ast}
\sim \{[cn]_{6_c}^{3_s}[\bar{c}\bar{n}]_{\bar{6}_c}^{3_s}\}_{1_c}^{1_s}\bigl|_{|\bm{I}| = 1}$
\end{tabular}\right.$   
& & &  \hspace{-5mm}
\begin{tabular}{l}
\vspace{-4mm}\\
 $\sim 3.3$ GeV \vspace{-1mm}\\
 ---------------\vspace{-2mm}\\
 \vspace{-2mm}\\
$\sim 3.9$ GeV
\end{tabular}
 & 
\begin{tabular}{l}
\hspace{0mm}$\hat{\delta}^c(3200)$\quad  ($\star$) \vspace{-0.5mm}\\
$\rightarrow \eta_c\pi,\,\psi\gamma$\vspace{-2.5mm}\\
--------------------\vspace{-2mm}\\
$\rightarrow D\bar{D}$ \\
$\rightarrow \eta_c\pi,\,\psi\rho$
\end{tabular}
\\
\cline{3-7}
\begin{tabular}{c}
\vspace{-13mm}\\
0\\ 
\end{tabular} & 
\begin{tabular}{c}
\vspace{-13mm}\\
0\\ 
\end{tabular} & & &
$\left.\begin{tabular}{l}
\hspace{1.5mm}$\hat{\sigma}^c
\sim \{[cn]_{\bar{3}_c}^{1_s}
         [\bar{c}\bar{n}]_{3_c}^{1_s}\}_{1_c}^{1_s}\bigl|_{|\bm{I}| = 0}$\vspace{-2.5mm}\\
------------------------------------\vspace{-2mm}\\
\hspace{0mm}$\hat{\sigma}^{c\ast}
\sim \{[cn]_{6_c}^{3_s}[\bar{c}\bar{n}]_{\bar{6}_c}^{3_s}\}_{1_c}^{1_s}\bigl|_{|\bm{I}| = 0}$
\end{tabular}\right.$     
&  \hspace{-5mm}
\begin{tabular}{l}
 $\sim 3.3$ GeV \vspace{-2mm}\\
 ---------------\vspace{-2mm}\\
$\sim 3.9$ GeV
\end{tabular}
& \hspace{-0mm}
\begin{tabular}{l}
\vspace{-4mm}\\
$\rightarrow \psi\gamma$\vspace{-2.5mm}\\
--------------------\vspace{-2mm}\\
$\rightarrow \eta_c\eta$
\end{tabular}
\\
\cline{3-7}
& 
& & & \hspace{-2mm}
$\left.\begin{tabular}{l}
\hspace{0.5mm}$\hat{\sigma}^{sc}\sim \{[cs]_{\bar{3}_c}^{1_s}
                                 [\bar{c}\bar{s}]_{3_c}^{1_s}\}_{1_c}^{1_s}$ 
\vspace{-2.5mm}\\
-----------------------------\vspace{-2mm}\\
\hspace{-1mm}$\hat{\sigma}^{sc\ast}\sim \{[cs]_{6_c}^{3_s}
                                                            [\bar{c}\bar{s}]_{\bar{6}_c}^{3_s}\}_{1_c}^{1_s}$
\end{tabular}\right.$   
&  \hspace{-5mm}
\begin{tabular}{l}
 $\sim 3.5$ GeV \vspace{-2mm}\\
 ---------------\vspace{-2mm}\\
$\sim 4.1$ GeV
\end{tabular}
& \hspace{-0mm}
\begin{tabular}{l}
\vspace{-4mm}\\
$\rightarrow \psi\gamma$\vspace{-2mm}\\
--------------------\vspace{-2mm}\\
$\rightarrow \eta_c\eta,\,\psi\phi$
\end{tabular}
\\
\hline
\end{tabular} \vspace{2mm}\\
%%%%%%%%%%%%%%%%%%%%%%%%%%%%%%%%%%%%%%%%%%%%%%%%%%%%%%%%%%%%%%%%%%%%%%%%
\hspace{50mm}
($\dagger$) Input data. \quad 
($\star$) A tiny $\eta\pi^0$ peak observed in $\gamma\gamma$ 
collisions at the Belle~\cite{hidden-charm-scalar-Belle}. 
\vspace{-2mm}
\end{table}
\end{center}
%%%%%%%%%%%%%%%%%%%%%%%%%%%%%%%%%%%%%%%%%%%%%%%%%%%%%%%%%%%%%%%%%%%%%%%% 

As discussed above, it is natural to consider that $D_{s0}^+(2317)$ and 
$\hat{\delta}^c(3200)$ are of $\bar{\bm 3}_c\times {\bm 3}_c$, while $X(3872)$ is of 
${\bm 6}_c\times \bar{\bm 6}_c$. 
In addition, the mass of $X^\pm(5568)$ has been reported as 
$m_{X(5568)} = 5567.8 \pm 2,9^{+0.9}_{-1.9}$ MeV~\cite{strange-bottom-D0}. 
If they truly exist and are charged components of iso-triplet bottom partners, 
{\small $X(5568)^{\pm,0} \sim \{[bn]_{\bar{3}_c}^{1_s}
               [\bar{s}\bar{n}]_{3_c}^{1_s}\}_{1_c}^{1_s}\bigr|_{|\bm{I}|=1}^{\pm,0},\,(n = u,d)$},  
of $D_{s0}^+(2317)$ with $\bar{\bm 3}_c\times {\bm 3}_c$, it will be expected that 
the mass difference $m_{X(5568)} - m_{D_{s0}^+(2317)}$ is not very far from 
$(m_{\eta_b} - m_{\eta_c})/2$ under the same quark counting as the above. 
Actually, their measured values~\cite{PDG16} are not very much different from each 
other, i.e., $m_{X(5568)} - m_{D_{s0}^+(2317)}\simeq 3250$ MeV and 
$(m_{\eta_b} - m_{\eta_c})/2\simeq 3210$ MeV, respectively, as expected. 
On the other hand, the mass of $X(3872)$ with ${\bm 6}_c\times \bar{\bm 6}_c$ is 
much higher than that of $\hat{\delta}^c$ with $\bar{\bm 3}_c\times {\bm 3}_c$ as 
discussed above, and therefore, it is expected, from the same quark counting, that 
the mass difference $m_{X(5568)} - m_{X(3872)}$ is much smaller than 
$m_{B_s^0} - m_{\eta_c}$. 
In fact, the measured value~\cite{PDG16} of the mass difference 
$(m_{X(5568)} - m_{X(3872)})_{\rm exp} \simeq 1.7$ GeV is much lower than 
$(m_{B_s^0} - m_{\eta_c})_{\rm exp}\simeq 2.4$ GeV, as expected above. 
If $D_{s0}^+(2317)$ and $X(3872)$ are truly of $\bar{\bm 3}_c\times {\bm 3}_c$ and 
of ${\bm 6}_c\times \bar{\bm 6}_c$, respectively, there would exist a rich spectrum of  
partners of $D_{s0}^+(2317)$ and $X(3872)$, i.e., scalar, axial-vector and tensor 
$\{[qq]_{6_c}^{3_s}[\bar{q}\bar{q}]_{\bar{6}}^{3_s}\}_{1_c}^{1_s,3_s,5_s}$ mesons with 
${\bm 6}_c\times \bar{\bm 6}_c$ and axial-vector 
%%%%%%%%%%%%%%%%%%%%%%%%%%%%%%%%%%%%%%%%%
$\{[qq]_{\bar{3}}^{1_s}(\bar{q}\bar{q})_{3_c}^{3_s} 
\oplus (qq)_{\bar{3}}^{3_s}[\bar{q}\bar{q}]_{3_c}^{1_s}\}_{1_c}^{3_s}$ 
%%%%%%%%%%%%%%%%%%%%%%%%%%%%%%%%%%%%%%%%%
with $\bar{\bm 3}_c\times {\bm 3}_c$. 
In this case, masses of 
$\{[qq]_{6_s}^{3_s}[\bar{q}\bar{q}]_{\bar{6}_c}^{3_s}\}_{1_c}^{1_s,3_s,5_s}$ 
are expected to be high enough for their hadronic decay modes (which are allowed 
under the OZI-rule~\cite{OZI}) to be open. 
In contrast, masses of 
%%%%%%%%%%%%%%%%%%%%%%%%%%%%%%%%%%%%%%%%%%%%%%%%%%
$\{[qq]_{\bar{3}}^{1_s}(\bar{q}\bar{q})_{3_c}^{3_s} 
\oplus (qq)_{\bar{3}}^{3_s}[\bar{q}\bar{q}]_{3_c}^{1_s}\}_{1_c}^{3_s}$ 
%%%%%%%%%%%%%%%%%%%%%%%%%%%%%%%%%%%%%%%%%%%%%%%%%%
will be much lower, so that most of them will decay through the electromagnetic 
interactions, except for some exceptinal cases. 

Open- and hidden-charm scalar mesons, $[cq]_{\bar{3}_c}^{1_s}[\bar{q}\bar{q}]_{3_c}^{1_s}$ 
and $[cq]_{\bar{3}_c}^{1_s}[\bar{c}\bar{q}]_{3_c}^{1_s}$, ($q = u,\,d,\,s$), respectively, in our 
tetra-quark model have been studied in \cite{D_{s0}-KT} and 
\cite{hidden-charm-scalar-KT}, where they were assumed to be of 
$\bar{\bm 3}_c \times {\bm 3}_c$. 
However, we now study tetra-quark states with ${\bm 6}_c\times \bar{\bm 6}_c$ 
in addition to $\bar{\bm 3}_c \times {\bm 3}_c$. 
Therefore, we put an asterisk ($\ast$) on each symbol of tetra-quark mesons with 
${\bm 6}_c\times \bar{\bm 6}_c$ to distinguish it from the corresponding one with 
$\bar{\bm 3}_c\times {\bm 3}_c$, for example, 
%%%%%%%%%%%%%%%%%%%%%%%%%%%%%%%%%%%%%%%%%%%%%%%%%%%%%%%%%%%%%%%%%%%%%%%%
{\small $\hat{F}_I^+
\sim \{[cn]_{\bar{3}_c}^{1_s}[\bar{s}\bar{n}]_{3_c}^{1_s}\}_{1_c}^{1_s}\bigl|_{|\bm{I}| = 1}^+$} 
with $\bar{\bm 3}_c\times {\bm 3}_c$ and 
{\small $\hat{F}_I^{*+}
\sim \{[cn]_{6_c}^{3_s}[\bar{s}\bar{n}]_{\bar{6}_c}^{3_s}\}_{1_c}^{1_s}\bigl|_{|\bm{I}| = 1}^+$} 
with ${\bm 6}_c\times \bar{\bm 6}_c$,    
%%%%%%%%%%%%%%%%%%%%%%%%%%%%%%%%%%%%%%%%%%%%%%%%%%%%%%%%%%%%%%%%%%%%%%%%
along with \cite{Jaffe}. 
We list this type of tetra-quark mesons 
%open- and hidden-charm scalar mesons  in the present model %are listed 
in Table I, in which $D_{s0}^+(2317)$ has been assigned to the iso-triplet $\hat{F}_I^+$, 
because it was observed in the $D_s^+\pi^0$ channel while no signal in the 
$D_s^{*+}\gamma$ channel~\cite{PDG16}. 
This fact means that its $D_s^+\pi^0$ decay is much stronger than the radiative 
$D_s^{*+}\gamma$ as expected from the hierarchy of hadron 
interactions~\cite{HT-isospin,KT-dilemma,Dalitz}, 
%%%%%%%%%%%%%%%%%%%%%%%%%%%%%%%%%%%%%%%%%%%%%%%%%%%%%%%%%%%%%%%%%%%%%%%%
$|${\em isospin conserving hadronic int.}$|\,\gg\,|${\em electromagnetic int.}$|\,\gg\,
|${\em isospin non-conserving hadronic int.}$|$. 
%%%%%%%%%%%%%%%%%%%%%%%%%%%%%%%%%%%%%%%%%%%%%%%%%%%%%%%%%%%%%%%%%%%%%%%%
In contrast, if it were an iso-singlet state as in \cite{PDG16}, it should decay 
dominantly through the electromagnetic interactions, because of the above hierarchy. 
In this case, it should be remembered that productions of the iso-singlet $\hat{F}_0^+$ 
in $e^+e^-$ annihilations are expected to be suppressed in comparison with the 
iso-triplet $\hat{F}_I^+$~\cite{production,KT-dilemma}. 
For these reasons, experiments should have detected $D_{s0}^+(2317)$ in the 
$D_s^{*+}\gamma$ channel of $B$ decays. 
Nevertheless, it was discovered in the $D_s^+\pi^0$ channel in inclusive $e^+e^-$ 
annihilations~\cite{D_{s0}-exp} and $B$ decays~\cite{Belle-D_{s0}}, while no signal in the 
$D_s^{*+}\gamma$ channel. 
This implies that the assignment of $D_{s0}^+(2317)$ to an iso-triplet state is quite 
natural.  
Possible decays of scalar tetra-quark mesons are tentatively listed in Table I, 
while only a part of them will be discussed later. 

As to axial-vector mesons, we study only hidden-charm flavorless 
$[cq](\bar{c}\bar{q}) \oplus (cq)[\bar{c}\bar{q}],\,(q = u,\,d,\,s)$ mesons in this note, 
because $X(3872)$ and $Z_c(3900)$ have been observed. 
(The other members will be studied elsewhere.) 
Here, ideally mixed hidden-charm $[cq](\bar{c}\bar{q})$ and $(cq)[\bar{c}\bar{q}]$ states 
belong to $\overline{\bm{60}}$- and $\bm{60}$-plets, respectively, of $SU_f(4)$, and  
two flavorless states in $\overline{\bm{60}}$- and $\bm{60}$-plets which consist of 
quarks with common flavors and have the same quantum numbers mix with each other 
to form $\mathcal{C}$-parity eigenstates,  
%%%%%%%%%%%%%%%%%%%%%%%%%%%%%%%%%%%%%%%%%%%%%%%%%%%%%%%%%%%%%%%%%%%%%%%%
\begin{eqnarray}
&&\hspace{-10mm}
\,X_I^{(\ast)0}(\pm) \,\,
= \frac{1}{\sqrt{2}}\{\,X_I^{\,(\ast)0}(\overline{\bm{60}}) \,
                                \pm \,\, X_I^{(\ast)\,0}({\bm{60}})\}\,
= \,\,\,\,\,\,\,\,\frac{1}{4}\,\,\,\Bigl\{\Bigl(
[uc]_{\bar{3}_c(6_c)}^{1_s(3_s)}(\bar{u}\bar{c})_{3_c(\bar{6}_c)}^{3_s(1_s)} 
- [dc]_{\bar{3}_c(6_c)}^{1_s(3_s)}(\bar{d}\bar{c})_{3_c(\bar{6}_c)}^{3_s(1_s)}\Bigr) 
\nonumber\\
&&\hspace{71mm} 
\pm \Bigl(
(uc)_{\bar{3}_c(6_c)}^{3_s(1_s)}[\bar{u}\bar{c}]_{3_c(\bar{6}_c)}^{1_s(3_s)} 
      - (dc)_{\bar{3}_c(6_c)}^{3_s(1_s)}[\bar{d}\bar{c}]_{3_c(\bar{6}_c)}^{1_s(3_s)}
\Bigr) \Bigr\},   
                                                                      \label{eq:decomp-X_I^ast(+-)}
\\
&&\hspace{-10mm}
X^{\,(\ast)}\,\,(\pm)\, = \frac{1}{\sqrt{2}}\{\,\,X^{\,(\ast)}\,(\overline{\bm{60}})\,\,
                                                              \pm \,\,\, X^{(\ast)}\,\,(\bm{60})\}\,
= \,\,\,\,\,\,\,\,\,\frac{1}{4}\,\,\,\Bigl\{\Bigl(
[uc]_{\bar{3}_c(6_c)}^{1_s(3_s)}(\bar{u}\bar{c})_{3_c(\bar{6}_c)}^{3_s(1_s)} 
+ [dc]_{\bar{3}_c(6_c)}^{1_s(3_s)}(\bar{d}\bar{c})_{3_c(\bar{6}_c)}^{3_s(1_s)}\Bigr) 
\nonumber\\
&&\hspace{70.5mm}  
\pm \,\Bigl((uc)_{\bar{3}_c(6_c)}^{3_s(1_s)}[\bar{u}\bar{c}]_{3_c(\bar{6}_c)}^{1_s(3_s)} 
        + (dc)_{\bar{3}_c(6_c)}^{3_s(1_s)}[\bar{d}\bar{c}]_{3_c(\bar{6}_c)}^{1_s(3_s)}
\Bigr)\Bigr\}, 
                                                                       \label{eq:decomp-X_I^{(ast)}(+-)}
\\
&&\hspace{-9.5mm}   
X^{s(\ast)}(\pm) 
= \frac{1}{\sqrt{2}}\{\,\,X^{s(\ast)}(\overline{\bm{60}}) \,\pm \,\, X^{s(\ast)}(\bm{60})\}
\,\,
= -\frac{1}{2\sqrt{2}}\Bigl\{\,\,\,\,
[sc]_{\bar{3}_c(6_c)}^{1_s(3_s)}(\bar{s}\bar{c})_{3_c(\bar{6}_c)}^{3_s(1_s)} \,
\pm\,\, (sc)_{\bar{3}_c(6_c)}^{3_s(1_s)}[\bar{s}\bar{c}]_{3_c(\bar{6}_c)}^{1_s(3_s)}\,\,
\Bigr\},
                                                                       \label{eq:decomp-X_I^{s(ast)}(+-)}
\end{eqnarray}
%%%%%%%%%%%%%%%%%%%%%%%%%%%%%%%%%%%%%%%%%%%%%%%%%%%%%%%%%%%%%%%%%%%%%%%%
where an asterisk $\ast$ has been put on each symbol of axial-vector states with 
${\bm 6}_c\times\bar{\bm 6}_c$, as in the scalar mesons, and the arguments $\pm$ 
denote the $\mathcal{C}$-parity eigenvalues. 
Although we assigned $X(3872)$ to $X(+)$ and studied its decay property in our 
earlier works~\cite{Terasaki-X,omega-rho-KT}, we now revise the assignment, i.e.,  
$X(3872) = X^{*}(+)$, as discussed before.  
In this case, the old assignment of $Z_c(3900)$ to $X_I(-)$~\cite{Z_c(3900)-KT} also 
should be revised, i.e., $Z_c(3900) = X_I^{*}(-)$. 
When $m_{X^*(+)} = m_{X(3872)} \simeq 3.9$ GeV is taken as the input data, the masses 
of $X_I^*(\pm)$, $X^*(-)$ and $X^{s*}(\pm)$ are very crudely estimated as 
$m_{X_I^*(\pm)} \simeq m_{X^*(-)} \simeq m_{X^*(+)} \simeq 3.9$ GeV and 
$m_{X^{s*}(\pm)} \simeq 4.1$ GeV by using the same quark counting, where the mass 
difference $\Delta_{sn} = m_s - m_n \simeq m_{D_s^+} - m_D\simeq 0.1$ GeV at the 
scale of charm meson mass has been taken. 

A gross feature of decay properties of tetra-quark mesons will be seen by 
decomposing each of them into a sum of products of $\{q\bar{q}\}$ pairs, and then, 
replacing a colorless spin-singlet {\small $\{q\bar{q}\}_{{1}_c}^{1_s}$} by a 
pseudoscalar meson with the corresponding flavor and iso-spin quantum number  
and a spin-triplet {\small $\{q\bar{q}\}_{{1}_c}^{3_s}$} by a vector meson as 
in \cite{Jaffe,HT-isospin}, where contributions of products of colored 
$\{q\bar{q}\}_{{8}_c}$ pairs will be dropped  in contrast to \cite{Jaffe}, and $J/\psi$ will 
be written as $\psi$. 
We list a part of results on decompositions of tetra-quark states under consideration, 
i.e., hidden-charm axial-vector $X^*(\pm)$, $X_I^{*0}(\pm)$ and $X^{s*}(\pm)$, and a 
hidden-charm scalar $\sigma^{sc*}$ with ${\bm 6}_c\times\bar{\bm 6}_c$ below. 
(As to decompositions of tetra-quark states with $\bar{\bm 3}_c\times{\bm 3}_c$, 
a part of them have been listed in our earier works  
\cite{HT-isospin,omega-rho-KT,Terasaki-X,Z_c(3900)-KT,hidden-strange}.) 
\begin{enumerate}
\item                  
Hidden-charm axial-vector tetra-quark mesons $X^{*}(\pm)$, $X_I^{*0}(\pm)$ and 
$X^{s*}(\pm)$ with ${\bm 6}_c\times\bar{\bm 6}_c$: 
%%%%%%%%%%%%%%%%%%%%%%%%%%%%%%%%%%%%%%%%%%%%%%%%%%%%%%%%%%%%%%%%%%%%%%%%
\begin{eqnarray}
&&\hspace{-15mm}
X^{*}(+) = \frac{1}{2\sqrt{6}}\Bigl\{2\bigl(\psi\omega - \omega\psi\bigr) 
                                 - \bigl[\bigl(D^0\bar{D}^{*0} + D^{*0}\bar{D}^0 \bigr) 
                                 - \bigl(\bar{D}^0D^{*0} + \bar{D}^{*0}D^0\bigr) \bigr] 
\nonumber\\
&& \hspace{45mm}
                                - \bigl[\bigl(D^+\bar{D}^{*-} + D^{*+}\bar{D}^-\bigr)  
                                - \bigl(\bar{D}^-D^{*+} + \bar{D}^{*-}D^+\bigr)\bigr] \Bigr\} 
                                + \cdots,
                                                                                     \label{eq:decomp-X*(+)}
\end{eqnarray}
%%%%%%%%%%%%%%%%%%%%%%%%%%%%%%%%%%%%%%%%%%%%%%%%%%%%%%%%%%%%%%%%%%%%%%%%
\begin{eqnarray}
&&\hspace{-15mm}
X^{*}(-) = \frac{1}{2\sqrt{3}}\Bigl\{- \bigl(\eta_c\omega - \omega\eta_c\bigr) 
                                  - \bigl(\psi\eta_0 - \eta_0\psi\bigr) 
\nonumber\\
&& \hspace{46mm}
                                 + \bigl[D^{*0}\bar{D}^{*0} - \bar{D}^{*0}D^{*0}\bigr] 
                                  - \bigl[D^{*+}\bar{D}^{*-} - \bar{D}^{*-}D^{*+}\bigr] \Bigr\} 
                                  + \cdots,
                                                                                     \label{eq:decomp-X*(-)}
\end{eqnarray}
%%%%%%%%%%%%%%%%%%%%%%%%%%%%%%%%%%%%%%%%%%%%%%%%%%%%%%%%%%%%%%%%%%%%%%%%
\begin{eqnarray}
&&\hspace{-14mm}
X_I^{*0}(+) = \frac{1}{2\sqrt{6}}\Bigl\{2\bigl(\psi\rho^0 - \rho^0\psi\bigr) 
- \bigl[\bigl(D^0\bar{D}^{*0} + D^{*0}\bar{D}^0 \bigr) 
- \bigl(\bar{D}^0D^{*0} + \bar{D}^{*0}D^0\bigr) \bigr] 
\nonumber\\
&& \hspace{45mm}
+\,\, \bigl[\bigl(D^+\bar{D}^{*-} + D^{*+}\bar{D}^-\bigr)  
- \bigl(\bar{D}^-D^{*+} + \bar{D}^{*-}D^+\bigr)\bigr] \Bigr\} + \cdots, 
                                                                                 \label{eq:decomp-X_I*(+)}
\\
&&\hspace{-15mm}
X_I^{*0}(-) = \frac{1}{2\sqrt{3}}\Bigl\{- \bigl(\eta_c\rho^0 - \rho^0\eta_c\bigr) 
                                  - \bigl(\psi\pi^0 - \pi^0\psi\bigr) 
\nonumber\\
&& \hspace{47mm}
                                - \bigl[D^{*0}\bar{D}^{*0} - \bar{D}^{*0}D^{*0}\bigr] 
                                  - \bigl[D^{*+}\bar{D}^{*-} - \bar{D}^{*-}D^{*+}\bigr] 
\Bigr\} + \cdots, 
                                                                                 \label{eq:decomp-X_I*(-)}
\\
&&\hspace{-15mm}
X^{s*}(+) = \frac{1}{2\sqrt{3}}\Bigl\{
\sqrt{2}\bigl[\psi\phi - \phi \psi \bigr] 
- \bigl[D_s^{*+}D_s^- - D_s^-D_s^{*+} \bigr] 
- \bigl[D_s^{+}D_s^{*-} - D_s^{*-}D_s^{+} \bigr]\Bigr\} + \cdots, 
                                                                                 \label{eq:decomp-Xs*(+)}\\
&&\hspace{-15mm}
X^{s*}(-) = \frac{1}{2\sqrt{3}}\Bigl\{
\sqrt{2}\bigl[D_s^{*+}D_s^{*-} - D_s^{*-}D_s^{*+}\bigr] 
- \bigl[\psi\eta_s - \eta_s \psi \bigr] 
- \bigl[\eta_c\phi - \phi\eta_c \bigr] \Bigr\} + \cdots, 
                                                                                 \label{eq:decomp-Xs*(-)}
\end{eqnarray}
%%%%%%%%%%%%%%%%%%%%%%%%%%%%%%%%%%%%%%%%%%%%%%%%%%%%%%%%%%%%%%%%%%%%%%%%
where $\eta_0$ and $\eta_s$ have been given by 
$\eta_0 = \eta\cos(\chi + \theta_P) + \eta'\sin(\chi + \theta_P)$ and 
$\eta_s = - \eta\sin(\chi + \theta_P) + \eta'\cos(\chi + \theta_P)$ 
under the ordinary $\eta\eta'$ mixing with the mixing angle $\theta_P$~\cite{PDG16},  
and $\chi$ satisfies $\cos(\chi) = \sqrt{1/3}$ and $\sin(\chi) = \sqrt{2/3}$. 
\item 
$\hat{\sigma}^{sc*}$ as a typical example of $[qq][\bar{q}\bar{q}]$ type of scalar 
tetra-quark mesons with ${\bm 6}_c\times\bar{\bm 6}_c$: 
%%%%%%%%%%%%%%%%%%%%%%%%%%%%%%%%%%%%%%%%%%%%%%%%%%%%%%%%%%%%%%%%%%%%%%%%
\begin{eqnarray}
&&\hspace{-10mm}
\hat{\sigma}^{sc*}
= \frac{1}{2\sqrt{{6}}}\Bigl\{
 \sqrt{3}(\eta_c\eta_s + \eta_s\eta_c) - \sqrt{3}(D_s^+{D}_s^- + {D}_s^-D_s^+) 
+ (\psi\phi + \phi \psi) - (D_s^{*+}{D}_s^{*-} + {D}_s^{*-}D_s^{*+})\Bigr\} + \cdots. 
                                                                     \label{eq:decomp-sigma^{sc-ast}}
\end{eqnarray}
%%%%%%%%%%%%%%%%%%%%%%%%%%%%%%%%%%%%%%%%%%%%%%%%%%%%%%%%%%%%%%%%%%%%%%%%
(Decompositions of the other members of $[qq][\bar{q}\bar{q}]$ and 
$[qq](\bar{q}\bar{q}) \oplus (qq)[\bar{q}\bar{q}]$ will be presented elsewhere.)
\end{enumerate}
%%%%%%%%%%%%%%%%%%%%%%%%%%%%%%%%%%%%%%%%%%%%%%%%%%%%%%%%%%%%%%%%

Although $X(3872)$ is now assigned to $X^{*}(+)$ with ${\bf 6}_c\times\bar{\bf 6}_c$, 
as discussed before, 
it is seen from Eq.~(\ref{eq:decomp-X*(+)}) that its possible decay modes are not 
drastically changed and the confirmed decay modes of $X(3872)$, i.e., 
%%%%%%%%%%%%%%%%%%%%%%%%%%%%%%%%%%%%%%%%%%%%%%%%%%%%%%%%%%%%%%%%%%%%%%%%
the isospin conserving 
$X(3872) \rightarrow D\bar{D}^{*}\oplus \bar{D}D^* \rightarrow D\bar{D}\pi^0$, 
$X(3872)\rightarrow \psi\omega \rightarrow \psi\pi^+\pi^-\pi^0$, 
the radiative 
$X(3872)\rightarrow \psi\omega\rightarrow \psi\gamma$ (under the vector meson 
dominance, VMD~\cite{VMD}) and 
the isospin non-conserving 
$X(3872)\rightarrow \psi\omega \rightarrow \psi\rho^0 \rightarrow \psi\pi^+\pi^-$ 
(through the $\omega\rho^0$ mixing),  
%%%%%%%%%%%%%%%%%%%%%%%%%%%%%%%%%%%%%%%%%%%%%%%%%%%%%%%%%%%%%%%%%%%%%%%%
are reproduced as in the old assignment~\cite{omega-rho-KT}, because their flavor 
wfs. are not changed, in contrast to its color and spin wfs. 
Regarding their rates, we do not study them at the present stage, because their 
experimental informations seem to be not sufficiently definite yet~\cite{PDG16}. 
(We need more definite informations for numerical analyses.)  
Here, it should be noted that a role of the $\omega\rho^0$ mixing in the isospin 
non-conserving $X(3872) \rightarrow \psi\rho^0 \rightarrow \psi\pi^+\pi^-$ 
decay~\cite{omega-rho-KT} should not be neglected, because it plays essential roles 
in the observed $\omega\rightarrow \pi^+\pi^-$ decay~\cite{PDG16} and in isospin 
non-conserving nuclear forces~\cite{nucl-force}. 
Under the revised assignment $X^*(+) = X(3872)$, Eq.~(\ref{eq:decomp-X_I*(+)}) 
implies that its iso-triplet partner $X_I^*(+)$ with the same $\mathcal{C}$-parity has 
an isospin conserving $X_I^*(+) \rightarrow \psi\rho \rightarrow \psi\pi\pi$ decay. 
Therefore, it is expected that its rate is very large and hence, its width is very 
broad, as in the old assignment~\cite{iso-triplet-X}, and hence, very high statistics will 
be needed to observe the iso-triplet partners of $X(3872)$ with the same  
$\mathcal{C}$ property. 
Although the assignment of $Z_c^{\pm,0}(3900)$ as iso-triplet parrtners of $X(3872)$ 
with opposite $\mathcal{C}$ property is also revised, i.e., 
$Z_c^{\pm,0}(3900) = X_I^{*\pm,0}(-)$, their OZI-rule-allowed decay modes are not 
drastically changed again. 
%%%%%%%%%%%%%%%%%%%%%%%%%%%%%%%%%%%%%%%%%%%%%%%%%%%%%%%%%%%%%%%%%%%%%%
On the other hand, $X(+)$ and $X_I(-)$ with $\bar{\bm 3}_c\times {\bm 3}_c$ (which 
were previously assigned to $X(3872)$ and $Z_c(3900)$) are now expected to have  
approximately degenerate masses much lower (by $\Delta_{\rm color}\simeq 0.6$ GeV) 
than the above $X^*(+)$ and $X_I^*(-)$, i.e., 
very crudely $m_{X(+)}\simeq m_{X_I(-)}\simeq 3.3$ GeV. 
As the result, $X(+)$ cannot have any OZI-rule-allowed decay modes but will decay 
dominantly through the electromagnetic interactions, while $X_I^{\pm,0}(-)$ might be 
able to decay exceptionally into $\psi\pi^{\pm,0}$ final states, if their mass is truly 
higher than the $\psi\pi$ threshold. 
%%%%%%%%%%%%%%%%%%%%%%%%%%%%%%%%%%%%%%%%%%%%%%%%%%%%%%%%%%%%%%%%%%%%%
(Although $X_I^{\pm}(\pm)$ and $X_I^{*\pm}(\pm)$ in the above are not 
$\mathcal{C}$-parity eigenstates, they are partners of the $\mathcal{C}$-parity 
eigenstates $X_I^{0}(\pm)$ and $X_I^{*0}(\pm)$, respectively, in each iso-triplet  
and satisfy $\mathcal{C}|X_I^{(*)\pm}(\pm)\rangle = \pm |X_I^{(*)\mp}(\pm)\rangle$ 
under the charge-conjugation.) 
%%%%%%%%%%%%%%%%%%%%%%%%%%%%%%%%%%%%%%%%%%%%%%%%%%%%%%%%%%%%%%%%%%%%%%
As seen in Eq.~(\ref{eq:decomp-X_I*(-)}), $X_I^{*0}(-)$ couples to $\psi\pi^{0}$ while it 
does not directly couple to $(D\bar{D}^{*})^0$ and $(\bar{D}D^{*})^0$ in our model, in 
contrast to a molecular model~\cite{Guo}. 
Therefore, the $D\bar{D}^{*}$ peak, $Z_c(3885)$, which has been observed at the 
BESIII~\cite{Z_c-DDast-exp} might not be identified with $Z_c(3900)$, i.e., it might be 
some kind of kinematical effect like a coupled-channel cusp~\cite{Swanson-cusp}. 

Now we return to scalar mesons listed in Table~I. 
As was seen in \cite{HT-isospin}, the isospin conserving 
$\hat{F}_I^+\rightarrow D_s^+\pi^0$ is the dominant decay of $\hat{F}_I^+$ which is 
assigned to $D_{s0}^+(2317)$~\cite{D_{s0}-KT}. 
Although this is allowed under the OZI rule, its rate can be small, because tetra-quark 
states have a variety of color and spin configurations, and as the result, 
the {\small ${\hat{F}_I^+\bar{D}_s^-\pi^0}$} coupling is suppressed at the scale of 
heavy meson mass, as discussed in \cite{HT-isospin,KT-Hadron-2003}. 
On the other hand, a dominant decay of its iso-singlet partner $\hat{F}_0^+$ will be 
$\hat{F}_0^+\rightarrow D_s^{*+}\omega\rightarrow D_s^{*+}\gamma$, when the VMD 
is accepted. 
Therefore, its detection in the $D_s^{*+}\gamma$ channel in $B$ decays is awaited, 
because its production in inclusive $e^+e^-$ annihilation will be suppressed as 
discussed before. 
In addition, it is expected that non-strange partners 
$\hat{D}\sim [cn][\bar{u}\bar{d}],\,(n =u,d)$ of $D_{s0}^+(2317)$ are narrow because of 
the same reason as the narrow width of $\hat{F}_I^+ =D_{s0}^+(2317)$. 
Their mass $m_{\hat{D}} \simeq 2.2$ GeV estimated by using the same quark counting 
as the above is lower by about 100 MeV than the mass 
$m_{D_0^*(2400)} = 2318\pm 29$ MeV~\cite{PDG16} of the conventional 
$D_0^{*}\sim$ {\small $^3P_0$} $\{c\bar{n}\},\,(n = u,d)$ mesons, and it is expected 
that $\hat{D}$ decays dominantly into the same $D\pi$ final states as the observed 
$D_0^{*}$.  
This implies that the tetra-quark $\hat{D}$ and the conventional $D_0^{*}$ co-exist in 
the observed broad $D\pi$ enhancement around 2.3 GeV and $D_0^{*}$ occupies 
its majour part, because its production rate is much higher and its width is much 
broader ($\Gamma_{D_0^*} = 267\pm 40$ MeV~\cite{PDG16}) than $\hat{D}$. 
(For more details, see \cite{KT-MacKellar}.)  
Therefore, $\hat{D}$ will be observed as a tiny peak on the lower tail of the broad 
$D\pi$ enhancement arising from $D_0^{*}$. 
The above argument might be compared with a recent discussion~\cite{Guo-two-pole} 
on two-pole structure of $D_0^{*}(2400)$. 
Returning to the charm-strange scalar sector, we expect existence of the 
conventional $^3P_0\,\{c\bar{s}\}$ scalar meson, $D_{s0}^{*+}$~\cite{KT-MacKellar}. 
Its mass is estimated as $m_{D_{s0}^{*+}}\simeq 2.4$ GeV by using the same quark 
counting with $\Delta_{sn}\simeq 0.1$ GeV as the above. 
This result, though still very crude, is not very far from our earier estimate by the 
QCD sum rule~\cite{HT-sum-rule}, and is high enough to decay into the $DK$ final state.  

As to hidden-charm scalar mesons, $\hat{\delta}^{c0}$ is interesting, because an 
indication of its candidate has been observed~\cite{hidden-charm-scalar-Belle}, as 
mentioned before. 
It will be narrow, for the same reason as the expected narrow widths of 
$D_{s0}^+(2317)$ and $\hat{D}$.  
In addition, it is suspected that the $\eta\pi^0$ peak will be tiny, because the 
$\eta\pi^0$ decay of $\hat{\delta}^{c0}$ is suppressed because of the OZI rule. 
This result is compatible with the observation of a tiny $\eta\pi^0$ peak around 3.2 
GeV at the Belle, mentioned before. 
On the other hand, open- and hidden-charm $[cq][\bar{q}\bar{q}]$ and 
$[cq][\bar{c}\bar{q}]$ with ${\bm 6}_c\times\bar{\bm 6}_c$ might not be narrow, 
because their masses will be high enough for various strong decay channels to be 
open.  

In summary, we have studied a part of open- and hidden-charm scalar tetra-quark 
mesons in addition to hidden-charm axial-vector tetra-quark ones, assigning 
%%%%%%%%%%%%%%%%%%%%%%%%%%%%%%%%%%%%%%%%%%%%%%%%%%%%%%%%%%%%%%%%%%%%%
$D_{s0}^+(2317)$ to 
{\small $\{[cn]_{\bar{3}_c}^{1_s}[\bar{s}\bar{n}]_{3_c}^{1_s}\}_{1_c}^{1_s}|_{|{\bm I}| = 1}^+$} 
and $X(3872)$ to 
{\small $\{[cn]_{6_c}^{3_s}(\bar{c}\bar{n})_{\bar{6}_c}^{1_s} 
+ (cn)_{6_c}^{1_s}[\bar{c}\bar{n}]_{\bar{6}_c}^{3_s}\}_{1_c}^{3_s}|_{|{\bm I}| = 0}$}.  
%%%%%%%%%%%%%%%%%%%%%%%%%%%%%%%%%%%%%%%%%%%%%%%%%%%%%%%%%%%%%%%%%%%%%
In this way, the measured large mass difference between $X(3872)$ and 
$D_{s0}^+(2317)$ has been naturally understood, in relation to the newly observed 
$X(5568)^\pm$. 
In addition, it has been discussed that our assignment of $D_{s0}^+(2317)$ to an 
iso-triplet state is quite natural. 
This assignment seems to be implicitly supported by the observation of charged 
$X(5568)^\pm$ as its iso-triplet bottom partners, that is, existence of $X(5568)$ as 
the bottom partner of $D_{s0}^+(2317)$ is quite natural in our model. 
Therefore, confirmation of existence of $X(5568)$ is awaited. 
In addition, this assignment expects existence of its neutral and doubly charged 
partners, $\hat{F}_I^0 = D_{s0}^0(2317)$ and $\hat{F}_I^{++} = D_{s0}^{++}(2317)$, while 
a recent experiment did not observe any indication of 
them~\cite{search-for-double-charge}. 
However, no signal of $D_{s0}^+(2317)$ in the $D_s^{*+}\gamma$ channel and no 
indication of $D_{s0}^0(2317)$ and $D_s^{++}(2317)$ are a serious (model-independent) 
dilemma in the $D_{s0}(2317)$ physics~\cite{KT-dilemma}. 
Therefore, observation of $D_{s0}^0(2317)$ and $D_s^{++}(2317)$ and/or a 
$D_s^{*+}\gamma$ peak around $m_{D_{s0}^+(2317)}$ is strongly desired. 
Regarding hidden-charm scalar mesons, an indication of a tiny $\eta\pi^0$ peak 
around 3.2 GeV (called as $\hat{\delta}^{c0}(3200)$ in this note) also has been 
understood easily in the present model, in contrast to the other existing 
models~\cite{D_{s0}-Maiani,hidden-charm-scalar-Oset}. 
Therefore, it is awaited that existence of $\hat{\delta}^{c0}(3200)$ is confirmed.  

Assignment of $X(3872)$ has been revised, i.e., $X(3872) = X^*(+)$ with 
${\bm 6}_c\times \bar{\bm 6}_c$ in this note. 
However, its decay property has not drastically been changed and the experimentally 
confirmed decay modes have been reproduced. 
Assignment of $Z_c^{\pm,0}(3900)$ as the iso-triplet partners of $X(3872)$ with an 
opposite $\mathcal{C}$ property also has been revised, i.e., $Z_c(3900) = X_I^{*}(-)$, 
and it has been seen that $Z_c(3900) = X_I^{*}(-)$ can decay into the $\psi\pi$, while 
it does not directly couple to $D\bar{D}^{*}$ and $\bar{D}D^{*}$ in our model. 
This might imply that the observed $D\bar{D}^{*}$ peak, $Z_c(3885)$, should not be 
identified with $Z_c(3900)$ but is some kind of kinematical effect. 

The hidden-charm and -strangeness scalar $\hat{\sigma}^{sc\ast}$ and axial-vector 
$X^{s\ast}(+)$ with ${\bm 6}_c\times\bar{\bm 6}_c$ are interesting in relation to 
recently observed $\psi\phi$ resonances~\cite{LHCb-psi-phi}, because each of them 
couples to a $\psi\phi$ state, as seen in Eqs.~(\ref{eq:decomp-Xs*(+)}) and 
(\ref{eq:decomp-sigma^{sc-ast}}). 
However, they will be studied elsewhere in future. 
Detailed analyses in scalar, axial-vector and tensor mesons,  
{\small $\{[qq]_{6_c}^{3_s}[\bar{q}\bar{q}]_{\bar{6}_c}^{3_s}\}_{1_c}^{1_s,3_s,5_s}$}, 
with ${\bm 6}_c\times \bar{\bm 6}_c$ and open-charm axial-vector mesons, 
{\small $\{[cq]_{\bar{3}_c(6_c)}^{1_s(3_s)}(\bar{q}\bar{q})_{3_c(\bar{6}_c)}^{3_s(1_s)} \oplus 
(cq)_{\bar{3}_c(6_c)}^{3_s(1_s)}[\bar{q}\bar{q}]_{3_c(\bar{6}_c)}^{1_s(3_s)}\}_{1_c}^{3_s}$}, 
with $\bar{\bm{3}_c}\times {\bm{3}_c}$ (and with ${\bm 6}_c\times \bar{\bm 6}_c$) are 
left intact as our future subjects. 

%%%%%%%%%%%%%%%%%%%%%%%%%%%%%%%%%%%%%%%%%%%%%%%%%%%%%%%%%%%%%%%%%%%%%%%%
\section*{Acknowledgments}    
The author would like to appreciate Professor T.~Hyodo for discussions and comments, 
and Professor H.~Kunitomo for carefull reading of the manuscript. 
He also would like to thank the Yukawa Institute for Theoretical Physics, Kyoto 
University, where this work was motivated during the workshop, YITP-T-15-08, 
on ``Exotic hadrons from high energy collision''.  
%%%%%%%%%%%%%%%%%%%%%%%%%%%%%%%%%%%%%%%%%%%%%%%%%%%%%%%%%%%%%%%%%%%%%%%

%%%%%%%%%%%%%%%%%%%%%%%%%%%%%%%%%

%\end{references}
%%%%%%%%%%%%%%%%%%%%%%%
%%%%%%%%%%%%%%%%%%%%%%%
\end{document}